\documentclass{ws-ijmpa}

\begin{document}

\markboth{A.\,V.~Kuznetsov, N.\,V.~Mikheev, A.\, M.~ Shitova}
{Ultra-High Energy Neutrino Dispersion in Plazma and Radiative Transition $\nu_L \to \nu_R + \gamma$}

%
\catchline{}{}{}{}{}
%

\title{ULTRA-HIGH ENERGY NEUTRINO DISPERSION IN PLASMA
\\ AND RADIATIVE TRANSITION $\nu_L \to \nu_R + \gamma$}


\author{\footnotesize A.\,V.~KUZNETSOV, N.\,V.~MIKHEEV, A.\,M.~SHITOVA}

\address{Division of Theoretical Physics, Department of Physics,\\
Yaroslavl State P.\,G.~Demidov University, Sovietskaya 14,\\
150000 Yaroslavl, Russian Federation\\
avkuzn@univ.uniyar.ac.ru, mikheev@univ.uniyar.ac.ru, pick@mail.ru}

\maketitle

\begin{history}
\received{Day Month Year}
\revised{Day Month Year}
\end{history}

\begin{abstract}
Qualitative analysis of additional energy of neutrino and
antineutrino in plasma is performed. A general expression for the 
neutrino self-energy operator is obtained in the case of
ultra-high energies when the local limit of the weak interaction is not valid.
The neutrino and antineutrino additional energy in plasma is calculated 
using the dependence of the $W$ and $Z$--boson propagators on the momentum transferred. 
The kinematical region for the neutrino radiative transition (the so-called ``neutrino spin light'') 
is established for some important astrophysical cases. 
For high energy neutrino and antineutrino, dominating transition channels in plasma, 
$\nu_e + e^+ \to W^+$, $\bar\nu_e + e^- \to W^-$ and $\bar\nu_{\ell} + \nu_{\ell} \to Z$, are indicated. 

\end{abstract}

\ccode{PACS numbers: 13.15.+g, 95.30.Cq}

\section{Introduction}
\label{sec:Introduction}

The neutrino physics development during the last decades, and
especially solving the Solar neutrino puzzle in the unique
experiment on the heavy-water detector at the Sudbery Neutrino
Observatory together with monitoring the Galaxy by the net of
neutrino detectors aimed to registrate a neutrino signal from the
expected galactic supernova explosion, brings to the fore the neutrino
physics in an active external medium.  The study of the
 external medium influence on the neutrino dispersive
properties is based on the analysis of the neutrino self-energy
operator.

The neutrino self-energy operator $\Sigma(p)\,$ can be defined in
terms of the invariant amplitude for the transition
$\nu\rightarrow\nu\,$, that is the neutrino coherent forward
scattering~\cite{Wolfenstein:1978}, by the relation:
\begin{equation}
\label{eq:Amplitude} {\cal M}
(\nu\rightarrow\nu)=-\left[\bar{\nu}(p)\Sigma(p)\nu(p) \right]
=-\text{Tr} \left[ \Sigma(p)\rho(p) \right],
\end{equation}
where $p^{\alpha}=(E,{\bf p})\,$ is the neutrino four-momentum,
$\rho(p)= \nu(p)\bar{\nu}(p)\,$ is the neutrino density matrix.
Effect of the external active medium on neutrino properties
specifies an appearance of the additional neutrino energy that can be
defined via the self-energy operator $\Sigma(p)\,$ as follows:
\begin{equation}
\label{eq:Add.Energy} \Delta
E=\frac{1}{2E}\text{Tr}[\Sigma(p)\rho(p)]{.}
\end{equation}
\noindent It should be mentioned that the medium influence on
neutrino properties is due primarily to the additional energy
acquired only by the left-handed neutrinos. The discovery of the neutrino
oscillations and hence of non-zero neutrino masses points to the
necessity of existence of the right-handed neutrinos, which are
sterile to the weak interactions and therefore are not acquiring
additional energy in medium. If a neutrino carries a magnetic moment,
there exists a possibility for interaction with photons leading to the 
neutrino spin flip. In this case the left-handed neutrino additional
energy appearance makes possible the neutrino radiative conversion:
\begin{equation}
\nu_L \to \nu_R + \gamma \,. \label{eq:nunugamma}
\end{equation}
 \noindent  This situation called the ``spin light of neutrino''
 ($SL \nu$), was first proposed and investigated in detail in an extended series of papers
  (see Ref.~\refcite{Studenikin:2006} and the papers cited therein). 
  However, in the analysis of this effect the authors
  missed the plasma influence on the photon dispersion.
  As it was shown in Refs.~\refcite{Kuznetsov:2006} and~\refcite{Kuznetsov:2007ijmpa}, 
  taking account of this influence
  makes the neutrino spin light process kinematically forbidden in almost all real astrophysical situations.
  In the latest publications (see e.g. Ref.~\refcite{Studenikin:2008}), a consideration of the $SL \nu\,$ process reduced to the limit of ultra-high neutrino energies.
Actually, in this case the dispersion properties of a photon can be
neglected. But the using of the weak interaction local limit would not be justified then.

There exists another physical possibility where the expression for the 
neutrino additional energy in plasma obtained in the local limit
of the weak interaction is insufficient. It occurs in the case of
nearly charged-symmetric plasma, e.g. in the conditions of the Early
Universe. In this case the local contribution to the neutrino
additional energy vanishes, and a part of the neutrino
additional energy caused by non-locality of the weak
interaction becomes essential. This contribution to the neutrino 
additional energy was investigated in Ref.~\refcite{Notzold:1988_NPB} 
(see also Refs.~\refcite{Langacker:1992} and~\refcite{Elmfors:1996}).

In the listed papers~\cite{Notzold:1988_NPB,Langacker:1992,Elmfors:1996,Kuznetsov:2007ijmpa}
the accounting of the non-local contribution to the neutrino
additional energy was made by the retention of the next term in
the expansion of the $W$-- and $Z$--boson propagators in the inverse powers of their masses. 
However in the limit of the ultra-high energies
this kind of expansion should be banned and therefore it is
necessary to use the exact expressions for the $W$-- and
$Z$--boson propagators. Analysis of the neutrino additional energy in a plasma in the limit of ultra-high energies, with taking account of the nonlocality of the weak interaction was made in a series of papers, Refs.~\refcite{Lunardini:2000,Lunardini:2001,Sahu:2008}, with
respect to the neutrino oscillations. In the present paper we
consider the neutrino self-energy operator in medium with taking into
account the dependence of the $W$ and $Z$--boson propagators on the momentum transferred, 
and we analyse its effects on the neutrino radiative conversion~\eqref{eq:nunugamma}.

\section{Neutrino Self-Energy Operator in~Medium}
\label{sec:Neu-self-en-operator}
%
Let us consider first the electron neutrino scattering on the
electron-positron component of plasma.

\begin{figure}
\begin{center}
\includegraphics*[width=0.5\textwidth]{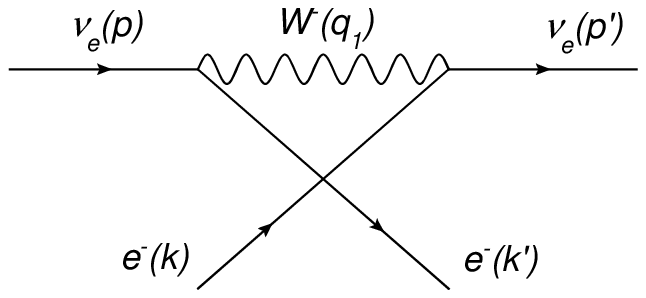}
\caption{The Feynman diagram for the neutrino-electron scattering through $W$--boson.}
\end{center}
\label{figure1}
\end{figure}

The Lagrangian of the interaction has the form:
\begin{equation}
\label{eq:lagrangian}
L=\frac{g}{2\sqrt{2}}\left[(\bar{e}\gamma_{\alpha}(1-\gamma_{5})\nu_{e})W^{\alpha}+
\text{h.c.} \right]{,}
\end{equation}
\noindent 
where $\gamma_5$ is used in notations of Ref.~\refcite{Bjorken:1964}, 
and leads to the invariant amplitude of the process:
\begin{eqnarray}
 && M_{\nu_{e}e^{-}\rightarrow
\nu_{e}e^{-}}=-\frac{G_{\mathrm{F}}}{\sqrt{2}}\left[\bar{e}(k')\gamma_{\alpha}(1-\gamma_{5})e(k)\right] \nonumber
\nonumber\\[3mm]
&&\times\left[\bar{\nu}_{e}(p')\gamma^{\alpha}(1-\gamma_{5})\nu_{e}(p)\right]\frac{1}{1-q_1^2/m_W^2}{\,,}\label{eq:amplitude2}
\end{eqnarray}
where we use the notation $q_{1}=k-p'\,$ for the
$W^{-}$--boson momentum (see Fig. 1). 
Here, the Fiertz transformation is performed, and the small term in the $W$--boson propagator 
of the order of $(m_e/m_W)^2$ is neglected.

\begin{figure}
\begin{center}
\includegraphics*[width=0.5\textwidth]{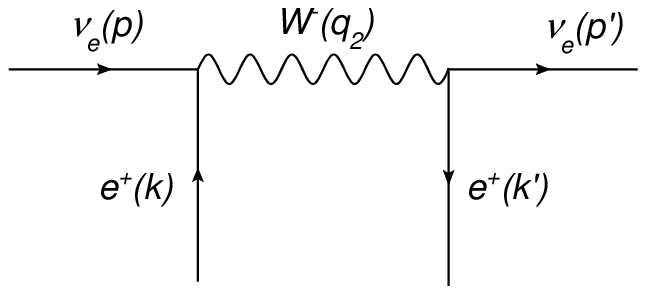}
\caption{The Feynman diagram for the neutrino-positron scattering
through $W$--boson.}
\end{center}
\label{fig2}
\end{figure}

The amplitude of the neutrino-positron scattering process can be written 
in the similar form (see Fig.~\ref{fig2}):
\begin{eqnarray}
&& M_{\nu_{e}e^{+}\rightarrow
\nu_{e}e^{+}}=\frac{G_{\mathrm{F}}}{\sqrt{2}}\left[\bar{e}(-k)\gamma_{\alpha}(1-\gamma_{5})e(-k')\right] \nonumber
\nonumber\\[3mm]
&&
\times\left[\bar{\nu}_{e}(p')\gamma^{\alpha}(1-\gamma_{5})\nu_{e}(p)\right]\frac{1}{1-q_2^2/m_W^2}{,}
\label{eq:Amplitude3}
\end{eqnarray}
where $W^{-}$--boson momentum is $q_{2}=-p-k$. Note that in
contrast to the $u$-channel process, described by the diagram in Fig.~1, 
the process in Fig.~\ref{fig2} is of the $s$-channel type. It means that in
this process, a resonance behavior of the $W$--boson propagator manifests itself. 
Taking account of this type of resonance is made 
by introducing a complex mass of $W$--boson, $m^{\ast}_{W}=m_{W}-\frac{1}{2} \, \mathrm{i} \,
\Gamma_{W}$, where $\Gamma_{W}$ is the total decay width of $W$--boson, $\Gamma_{W} \simeq 2.1\,$GeV.

Because of the $t\,$--channel behavior of the neutrino-electron
and neutrino-positron scattering diagrams for neutrinos of all
flavors through $Z$--boson, and keeping in mind that the
forward scattering is considered, i.e. the scattering with zero-momentum
transfer, one concludes that the contribution to the energy from these subprocesses 
is described by the local limit of the weak interaction. 

The total contribution to the neutrino self-energy
operator for $\ell\,$--flavor neutrino from the neutrino scattering processes on plasma electrons and
positrons can be represented in the form:
\begin{eqnarray}
\Sigma^{\nu_{\ell}}_{(e^- e^+)} (p) = \sqrt{2}G_{\mathrm{F}} \left[C_{V}
(u\gamma)\gamma_{L}(N_{e}-\bar{N}_{e}) + \delta_{\ell e}\gamma^{\alpha}\gamma_{L}
(j^{-}_{\alpha}-j^{+}_{\alpha})\right]{,}
\label{eq:self-en-operator2}
\end{eqnarray}
where $\gamma_{L} = (1-\gamma_{5})/2$, 
$N_{e},\bar{N}_{e}=2(2\pi)^{-3}\int
\mathrm{d}^{3}k\left(\exp \left({(\varepsilon\mp
 \mu)/T}\right)+1\right)^{-1}\,$ are the electron and positron densities respectively,
 and we use the notation
\begin{equation}
\label{eq:j} j^{\mp}_{\alpha}= 2\int\frac{\mathrm{d}^3
k}{(2\pi)^3}\frac{k_{\alpha}}{\varepsilon}\left(\mathrm{e}^{\frac{\varepsilon\mp
 \mu}{T}}+1\right)^{-1}\left(1 \pm \frac{2(kp)}{m_{W}^2}\right)^{-1}{.}
\end{equation}
The constant $C_{V}$ in Eq.~\eqref{eq:self-en-operator2} comes from the electron $Z$--current,
 $C_{V}=-1/2+2\,\sin^2\,\theta_{\mathrm{W}}$,
  where $\theta_{\mathrm{W}}\,$  is the Weinberg angle.

In accordance with Eq.~\eqref{eq:Add.Energy}, the neutrino
$\nu_{\ell}$ additional energy in the electron and positron medium
takes the form:
\begin{eqnarray}
\Delta E^{\nu_{\ell}}_{(e^- e^+)} = \sqrt{2} G_{\mathrm{F}} \left[C_{V}
(N_{e}-\bar{N}_{e})+ \, \delta_{\ell e}
\left(F_{1}(\mu_{e},m_{W})-F_{2}(-\mu_{e},m_{W})\right)\right]{,}\label{eq:9}
\end{eqnarray}
where we introduce the functions
\begin{eqnarray}
F_{1,2}(\mu,m) = \frac{2}{(2\pi)^3 E}\int
\frac{\mathrm{d}^3
k}{\varepsilon}\left(\mathrm{e}^{\frac{\varepsilon-\mu}{T}}+1\right)^{-1}
\frac{(pk)}{\left(1\pm \frac{2(pk)}{m^2}\right)} {.} 
\label{eq:10}
\end{eqnarray}

In order to obtain the antineutrino additional energy in the same
medium, one has to make the replacement $\mu_{e}\rightarrow -\mu_{e}$
in the right-hand side of Eq.~\eqref{eq:9}. 
In the first term with the difference of the electron and positron densities 
it simply means a change of sign.

In the analysis of the neutrino dispersion in active
astrophysical medium in a general case, the presence of the other plasma components, 
protons and neutrons, must be considered. 
In a dense plasma of the supernova core the donation from thermal neutrinos 
that can be considered to be approximately in equilibrium, can also be
significant. The two of the four Feynman diagrams for the 
neutrino-neutrino interaction contain a contribution from the
non-locality of weak interaction. 

A complete formula for the $\nu_{\ell}\,$ neutrino and
$\bar\nu_{\ell}\,$ antineutrino additional energy can be written
in the following way:
\begin{eqnarray}
 &&
 \Delta E^{\nu_{\ell},\bar\nu_{\ell}} = \sqrt{2} G_{\mathrm{F}}
\bigg\{ \mp \frac{1}{2}(N_{n}-\bar{N}_{n}) \pm
(N_{\nu_{e}}-\bar{N}_{\nu_{e}}) 
\nonumber\\[3mm]
&& \pm (N_{\nu_{\mu}}-\bar{N}_{\nu_{\mu}}) \pm
(N_{\nu_{\tau}}-\bar{N}_{\nu_{\tau}}) 
\nonumber\\[3mm]
&& + \delta_{\ell e}\left[F_{1}(\pm \mu_{e},m_{W})-F_{2}(\mp
\mu_{e},m_{W})\right] 
\nonumber\\[3mm]
&& + \, \frac{1}{2}\left[F_{1}( \pm \mu_{\nu_{\ell}},m_{Z})-F_{2}(
\mp \mu_{\nu_{\ell}},m_{Z})\right]\bigg\} {.}
\label{eq:11}
\end{eqnarray}
In this expression, $N_{n},N_{\nu_{\ell}}\,$ are the neutron and
neutrino densities and $\bar{N}_{n},\bar{N}_{\nu_{\ell}}\,$
are the densities of the corresponding antiparticles.
Electron and proton densities are cancelled in Eq.~\eqref{eq:11} because of plasma
electroneutrality. Note that in both functions $F_2$ there
exists the mentioned above resonance behavior, which can be
accounted by the introduction of complex masses of $W$-- and
$Z$--bosons, $m^{\ast}_{W,Z}=m_{W,Z}-\frac{1}{2} \, \mathrm{i}
\, \Gamma_{W,Z}$, where the total decay width of the $Z$-- boson
is $\Gamma_{Z}\simeq 2.5\,$GeV.

Tending formally $m_{W}$ and $m_{Z}$ in Eq.~\eqref{eq:11} to infinity, one obtains the neutrino additional energy in
the local limit of weak interaction, the so-called Wolfenstein 
energy~\cite{Wolfenstein:1978}. The additional energy obtained by this way is
inapplicable in the case of charge-symmetric plasma, e.g. in the Early
Universe. One has to take into account the additional
contribution to the neutrino energy caused by the non-locality
of weak interaction. This kind of energy was investigated in Refs.~\refcite{Notzold:1988_NPB,Langacker:1992,Elmfors:1996}. 
The non-local correction to the Wolfenstein energy was taken in the form of
the next terms in the expansion of the $W$-- and $Z$--boson propagators by the
inverse powers of their masses $m^{-2}_{W,Z}$. So, the first
correction can be obtained 
from Eq.~\eqref{eq:11}, if one retains the first term in the
expansion of the functions $F_{1,2}$ by $m^{-2}$. This
correction has the form:
\begin{eqnarray}
\Delta^{(1)}E^{\nu_{\ell}} = -\frac{16
G_{\mathrm{F}} E}{3\sqrt{2}}\left( \frac{\langle E_{\nu_{\ell}}
\rangle N_{\nu_{\ell}}+\langle E_{\bar\nu_{\ell}} \rangle
\bar{N}_{\nu_{\ell}} }{m_{Z}^{2}} + \, \delta_{\ell e} \, \frac{\langle E_{e} \rangle
N_{e} + \langle E_{\bar e} \rangle \bar{N}_{e}}{m_{W}^{2}} \right)
{,} \label{eq:14}
\end{eqnarray}
which coinsides with the result of Ref.~\refcite{Notzold:1988_NPB}. 
Here, $\langle E_{\nu_{\ell}} \rangle, \langle
E_{\bar\nu_{l}} \rangle, \langle E_{e} \rangle, \langle E_{\bar e}
\rangle $ are the average energies of plasma neutrinos,
antineutrinos, electrons and positrons respectively. However,
the correction of the type of Eq.~\eqref{eq:14} can be insufficient in the
case of ultra-high neutrino or antineutrino energies.
 That is why it is interesting to obtain the neutrino self-energy
 operator with using the dependence of the propagators of gauge bosons on the momentum transferred. 

\section{Kinematically Possible Regions For\\ Neutrino Radiative Conversion in Plasma}
\label{sec:kinematically-possible-regions}

 In the analysis of a kinematical possibility for the neutrino radiative
conversion~\eqref{eq:nunugamma} there can be essential three physical parameters,
 namely: the energy of the initial neutrino
$E$, the neutrino additional energy in plasma $\Delta E$ and the effective
photon (plasmon) mass $m_{\gamma}$. The existence of the
neutrino additional energy leads to the appearance of the
effective squared mass $m_{L}^2\,$ of the left-handed neutrinos:
\begin{equation}
\label{eq:30} m_{L}^2={\cal P}^2=(E+\Delta E)^2-{\bf p}^{\,2}{,}
\end{equation}
where ${\cal P}\,$ is the neutrino four-momentum in plasma in the
plasma rest frame, while $(E,{\bf p})$  should denote the neutrino
4-momentum in vacuum, $E = \sqrt{{\bf p}^2 + m_\nu^2} \simeq |{\bf p}|$. 
Hereafter we neglect the vacuum neutrino mass $m_\nu$, because in real 
astrophysical situations where $\Delta E$ could play any role, $m_\nu$ is less 
than $\Delta E$ and much less than $m_{\gamma}$. 

A condition for the kinematic opening of the process~\eqref{eq:nunugamma} has the form of the following
inequality:~\cite{Kuznetsov:2006,Kuznetsov:2007ijmpa}
\begin{equation}
m_L^2 \simeq 2 \, E \, \Delta E > m_\gamma^2 \,. 
\label{eq:30a}
\end{equation}
Because of the dependence of the neutrino additional energy $\Delta E$ on the 
neutrino energy $E$, see Eqs.~\eqref{eq:10}, \eqref{eq:11}, the inequality~\eqref{eq:30a} 
could be non-trivial. Let us consider it for different astrophysical situations.

\subsection{Nonrelativistic Cold Plasma}
\label{sec:nonrel-cold-plasma}

Let us consider first the high-energy neutrino propagation through the ``cold'' plasma of
the Sun or of red giants, where the temperature is $T \sim (10^7-10^8)
\, \mathrm{K} \sim (10^{-3}-10^{-2}) \, m_e$, and the electron density
is  $N_e \sim 10^{26}\, \text{cm}^{-3}$. The effective
plasmon mass in these conditions takes the form:
 $m_{\gamma}=\sqrt{4\pi \alpha N_{e}/m_{e}}$.
In this situation we can assume electrons to be nonrelativistic,
 $k^{\mu}\simeq(m_{e},{\bf 0})$, so that $(p-k)^2\simeq-2m_{e}E$.
The stellar substance is transparent for the neutrino radiation, thus the
contribution for the neutrino additional energy from thermal neutrinos can be
neglected.

In these conditions, the electron gas can be considered as degenerate with a good accuracy. As a result, an integration in the functions
$F_{1,2}(\mu_{e},m_{W})\,$, see Eq.~\eqref{eq:10}, reduces to a
computation of the electron density, $N_{e}=Y_{e} N_{B}$, where
$Y_{e}\,$ is the electron fraction, and $N_{B}\,$ is the baryon density. 
The additional energy for a neutrino and antineutrino  is
\begin{equation}
\label{eq:18} \Delta E^{\nu_{\ell},{\bar\nu}_{\ell}} =
\sqrt{2}G_{\mathrm{F}}N_{B} \left(\pm\frac{\delta_{\ell e}Y_{e}}{1
\pm  2m_{e}E (m_{W})^{-2}} \mp\frac{1}{2}(1-Y_{e})\right){.}
\end{equation}
%
Insertion of the complex $W$--boson mass, $m_{W}^*$ is essential for the 
electron antineutrino only, to avoid a pole of $\Delta E$ at $E = m_W^2/(2m_e)$. 
The analysis of the threshold inequality~\eqref{eq:30a} for the 
electron neutrino reduces, in view of \eqref{eq:18}, to the
investigation of the positiveness of the square trinomial with
respect to the energy $E$. Assuming that inside of the Sun
$Y_{e}\simeq 0.6$, we conclude that the inequality~\eqref{eq:30a} is
not satisfied for any neutrino energies.

In the earlier papers~\refcite{Kuznetsov:2006,Kuznetsov:2007ijmpa}
where the local limit of the weak interaction was used, it was concluded 
that the neutrino radiative conversion in the considered conditions is possible 
for neutrino energies $E$ greater than threshold energy $E_0 \simeq 10^7\,$ GeV. 
One can see that taking account of the non-locality of the weak interaction leads to the total
closing of the effect for the electron neutrino in the nonrelativistic ``cold'' plasma.

Consider now the possibilities for a trueness of the inequality~\eqref{eq:30a} 
in the same conditions for other neutrino flavors.
Note that the question about any observational realization of this
process remains open.

The analysis of the inequality~\eqref{eq:30a} for the electron
antineutrino, where a real part of $\Delta E$ should be taken, 
shows that the radiative neutrino conversion is
possible for antineutrino energies greater than the threshold energy
value, $E > E_0 \simeq 0.6 \times 10^7\,$ GeV.

An imaginary part of $\Delta E^{{\bar\nu}_e}$ deserves a separate analysis. 
In general, the non-zero imaginary part of a self energy means an instability of a particle. 
In the considered case it means that the electron antineutrino is unstable with respect 
to the process $\bar\nu_e + e^- \to W^-$ on plasma electrons. Using the formula for the width of the process: 
\begin{equation}
\label{eq:width_via_DE} 
w = - 2 \, \text{Im} \, \Delta E \,,
\end{equation}
one obtains from Eq.~\eqref{eq:18}: 
\begin{equation}
\label{eq:width_Sun} 
w (\bar\nu_e + e^- \to W^-) = 2 \sqrt{2} \, G_{\mathrm{F}} N_e E_0 \, \frac{\Gamma_W E_0/m_W}
{(E - E_0)^2 + (\Gamma_W E_0/m_W)^2} \,,
\end{equation}
where $E_0 = m_W^2/(2 m_e)$. 
Evaluation of a mean free path with respect to this process, $\lambda = 1/w$, 
for $N_e \sim 10^{26} \, \text{cm}^{-3}$, $E \sim 10^7\,$GeV provides $\lambda \sim 100\,$km, 
while in the maximum of the width defined by Eq.~\eqref{eq:width_Sun} at $E = E_0$ one obtains $\lambda \sim 200\,$m. 
It is obvious, that the process $\bar\nu_e + e^- \to W^-$ dominates the radiative 
neutrino conversion, see Refs.~\refcite{Kuznetsov:2006,Kuznetsov:2007ijmpa}.  
If one formally takes the limit $\Gamma_W \to 0$ in Eq.~\eqref{eq:width_Sun} to obtain:
\begin{equation}
\label{eq:width_Sun2} 
w (\bar\nu_e + e^- \to W^-) = 2 \sqrt{2} \, \pi \, G_{\mathrm{F}} N_e E_0 \, \delta (E - E_0) \,.
\end{equation}
It coinsides with the result of a direct calculation of the $W$--boson production by $\bar\nu_e$ 
scattered off nonrelativistic electron gas, without taking account of the instability of 
the $W$--boson. 

The interaction of the $\mu$- and $\tau$-neutrinos with medium occurs only through 
the $Z$--boson exchange with the zero momentum transfer and, as it
was pointed above, it is completely described by the local limit
of the weak interaction. As it can be seen from Eq.~\eqref{eq:18}, 
the $\nu_\mu, \; \nu_\tau$ additional energy is
negative, consequently the neutrino radiative conversion process is
closed for these neutrino flavors.

In turn, the antineutrino $\bar\nu_\mu\,$ and $\bar\nu_\tau\,$ additional energy is
positive. To estimate the border of the kinematically possible
region for the $SL \nu$ process in this case one can use a simple inequality:
\begin{equation}
\label{eq:32} E > E_0 = 4 \, \sin^2\,\theta_{W} \,
\frac{Y_{e}}{1-Y_{e}} \, \frac{m_{W}^2}{m_{e}}{.}
\end{equation}
For $Y_{e}\simeq 0.6$, the process is kinematically opened for
$\mu\,$-- and $\tau\,$--antineutrino energies greater than
$E_0\simeq 2 \times 10^{7}\,$GeV.

\subsection{Neutron Stars}
\label{sec:neutron-stars}

The substance of a neutron star is transparent for the neutrino radiation, 
as in the previous case. Electrons in extremely dense 
neutron stars are ultra-relativistic, therefore $\mu_{e}\simeq
p_{\mathrm{F}} \simeq 120 \, (N_e/(0.05 \, N_0))^{1/3}\,$ MeV,
where $p_{\mathrm{F}}\,$ is the electron Fermi momentum, and
$N_{0}=0.16\,$ Fm$^{-3}$\, is the typical nuclear density~\cite{Potehin:2010}. 
Due to the modern estimations, the
temperature inside neutron stars does not exceed a part of MeV, so the 
electron gas can be considered to be degenerate and an approximation of the zero 
temperature can be used. In this
case the electron density is $N_{e}=\mu_{e}^3/(3\pi^2)\,$ and the 
square effective plasmon mass is
 $m_{\gamma}^2=2\alpha\mu_e^2/ \pi$.

The additional energy for an electron neutrino under such conditions takes the following
form:
\begin{equation}
 \Delta
E^{\nu_e}=\sqrt{2}G_{\mathrm{F}}\left( - \frac{1}{2}\, (1 - Y_e) \, N_B +\frac{1}{2\pi^2}A(E,\mu_e)\right){,}
\label{eq:19}
\end{equation}
\begin{eqnarray}
&& A(E,\mu_e)=\frac{1}{16 E^3} \bigg[4 Em_{W}^2
\mu_{e}(m_{W}^2+2E\mu_e)
\nonumber\\[3mm]
&& -(m_{W}^6+4E\mu_{e}m_{W}^4)\ln \left(1+ \frac{4E\mu_e}{m_W^2}
\right) \bigg] {.}
\label{eq:20}
\end{eqnarray}

The analysis of the threshold inequality~\eqref{eq:30a} with
taking account of Eqs.~\eqref{eq:19}, \eqref{eq:20}
indicates that the $SL\nu\,$ process for the electron neutrino is
forbidden in the conditions of a neutron star.

The similar analysis can be held for the antineutrino. The additional
energy in this case is
\begin{equation}
 \Delta
E^{\bar{\nu}_{e}}=\sqrt{2}G_{\mathrm{F}}\left( \frac{1}{2}\, (1 - Y_e) \, N_B -\frac{1}{2\pi^2}\bar{A}(E,\mu_e)\right){,}\label{eq:21}
\end{equation}
\begin{equation}
 \bar{A}(E,\mu_e)=\int\limits_{0}^{\mu_{e}}k^2 \mathrm{d} k
\int\limits_{-1}^{1}\frac{(1-x) \mathrm{d}
x}{1 - \frac{2E(1-x) k}{m_{W}^2} - \mathrm{i} \frac{\Gamma_W}{m_W} } \,{.}
\label{eq:21a}
\end{equation}
This integral can be easily calculated analytically but the final
expression is too cumbersome. From the analysis of the kinematically
possible region~\eqref{eq:30a}, where a real part of $\Delta E$ should be taken, 
we can conclude that the radiative conversion process~\eqref{eq:nunugamma} 
is permitted for the electron antineutrino for energies greater than 
the threshold value $E_0 \simeq 8 \times 10^4\,$GeV, for $Y_e \simeq 0.1$, $N_B \simeq
10^{37} \, \text{cm}^{-3}$.

A comparison of these conclusions with the results of 
Refs.~\refcite{Kuznetsov:2006,Kuznetsov:2007ijmpa} shows that
taking account of the non-locality of the weak interaction
does not lead to any qualitative changes of the conclusions on 
kinematical possibilities of the radiative conversion for the electron
neutrino and antineutrino in the conditions of a neutron star.

Again, as in the considered case of ``cold'' plasma, an imaginary part 
of $\Delta E^{{\bar\nu}_e}$ means an instability of 
the electron antineutrino with respect to the process $\bar\nu_e + e^- \to W^-$ 
on plasma electrons. A width of the process can be obtained from Eqs.~\eqref{eq:width_via_DE}, \eqref{eq:21}, \eqref{eq:21a}, but in a general case the expression is rather cumbersome. 
It is esssentially simplified for high neutrino energies, $E \gg m_W\, \Gamma_W/\mu_e$, 
taking the form: 
\begin{equation}
\label{eq:width_NS} 
w (\bar\nu_e + e^- \to W^-) = \frac{G_{\mathrm{F}} m_W^4 \mu_e}{2 \sqrt{2} \, \pi \, E^2} 
\left( 1 - \frac{m_W^2}{4 \mu_e E} \right) \, \theta \left( E - \frac{m_W^2}{4 \mu_e} \right) \,.
\end{equation}
Evaluation of a mean free path with respect to this process  
for $\mu_{e} \simeq 120\,$MeV, $E \simeq 5 \times 10^4\,$GeV 
provides $\lambda \sim 10^{-5}\,$cm. 
Domination of the process $\bar\nu_e + e^- \to W^-$ over the radiative 
neutrino conversion in the neutron star conditions is undoubted, 
see Refs.~\refcite{Kuznetsov:2006,Kuznetsov:2007ijmpa}.  

For $\mu\,$--, $\tau\,$--neutrino and antineutrino, as well as in
the case of ``cold'' plasma, it is correct to use the local limit of the weak
interaction. Substituting the additional energy for $\ell = \mu,
\tau$
\begin{eqnarray}
 \Delta E^{\nu_{\ell},{\bar\nu}_{\ell}} = \mp
\frac{G_{\mathrm{F}}}{\sqrt{2}} \, (1 - Y_e) \, N_B
\,,\label{Addl_E}
\end{eqnarray}
and the plasmon mass in the case of a cold degenerate plasma
\begin{eqnarray}
\label{m_gamma} m_\gamma = \left(\frac{2 \, \alpha}{\pi}
\right)^{1/2} \left(3\, \pi^2 \, Y_e \, N_B \right)^{1/3}
\end{eqnarray}
into the threshold inequality~\eqref{eq:30a}, we come to the
conclusion that for $\nu_\mu, \; \nu_\tau$ the radiative conversion
process~\eqref{eq:nunugamma} is forbidden. For ${\bar\nu}_{\mu}, \, {\bar\nu}_{\tau}$
the process is kinematically permitted for the energies greater
than
\begin{eqnarray}
\label{NS_thr} E > E_0 = \frac{2 \, \sin^2\,\theta_W}{1-Y_e}
\left(\frac{3 \, Y_e}{\pi} \right)^{2/3} \frac{m_W^2}{N_B^{1/3}} .
\end{eqnarray}
Using for estimation the values $Y_e \simeq 0.1$, $N_B \simeq
10^{37} \, \text{cm}^{-3}$, we obtain $E_0 \simeq 2 \times 10^4\,$
GeV.

\subsection{Hot Plasma of a Supernova Core}
\label{sec:hotplasma-supernova}

In this case one needs to use the general expression for the neutrino
$\nu_{\ell}$ and antineutrino $\bar\nu_{\ell}$ additional energy
\eqref{eq:11}  with taking account of the scattering on all plasma
components. The additional energy can be written as:
\begin{eqnarray}
&& \Delta E^{\nu_{\ell},\bar\nu_{\ell}} =
\sqrt{2} G_{\mathrm{F}} \bigg\{ \mp \frac{1}{2}(N_{n}-\bar{N}_{n})
\pm
(N_{\nu_{e}}-\bar{N}_{\nu_{e}}) 
\nonumber\\[3mm]
&&\pm (N_{\nu_{\mu}}-\bar{N}_{\nu_{\mu}}) \pm
(N_{\nu_{\tau}}-\bar{N}_{\nu_{\tau}})  
\nonumber\\[3mm]
&&+ \frac{T^3}{2\pi^2} \bigg[ \delta_{\ell e}\big(B
(\pm\mu_{e},m_{W},T) - B (\pm\mu_{e},m_{W},-T)\big) 
\nonumber\\[3mm]
&& + \frac{1}{2} \big( B (\pm\mu_{\nu_{\ell}},m_{Z},T)- B
(\pm\mu_{\nu_{\ell}},m_{Z},-T)\big) \bigg] \bigg\} {,}
\label{eq:addl_E_SN} 
\end{eqnarray}
where we use the notation
\begin{eqnarray}
\label{eq:24} 
B (\mu,m,T)=-\frac{m^2}{ET}\bigg[
\text{Li}_{2}\left(\mathrm{e}^{-\frac{\mu}{T}}\right) 
+ a\int\limits_{0}^{\infty}\frac{\mathrm{d} y}{\exp
\left(y-\mu/T\right)+1} \ln \left| 1+\frac{y}{a} \right| \bigg]{.}
\end{eqnarray}
Here, $\text{Li}_{2} (z)$ is the Euler dilogarithm, and $a\,$ is the 
dimensionless parameter, $a=m^2/4ET$.

In the limit $m_{W}^2 \gg 4ET$, that is $a \gg 1$,
assuming that plasma is not degenerate ($\mu\sim T$), the integral
in Eq.~\eqref{eq:24} can be represented as the series expansion that can
be calculated analytically:
\begin{eqnarray}
\label{eq:25} 
\int\limits_{0}^{\infty}\frac{dy}{e^{-\mu/T}e^{y}+1}\text{ln}\left(1+\frac{y}{a}\right) &=&
e^{-\mu/T}\int\limits_{0}^{\infty}\frac{ydy}{e^{-\mu/T}e^{y}+1}-\frac{1}{2}e^{-2\mu/T}
\int\limits_{0}^{\infty}\frac{y^2dy}{e^{-\mu/T}e^{y}+1}
\nonumber\\[3mm]
& +&
\frac{1}{3}e^{-3\mu/T}\int\limits_{0}^{\infty}\frac{y^3dy}{e^{-\mu/T}e^{y}+1}-\ldots
\end{eqnarray}

Taking into account that the Fermi integrals are expressed in terms of polylogarithms:
\begin{equation}
\label{eq:26} 
\int\limits_{0}^{\infty}\frac{y^n
dy}{e^{-\mu/T}e^{y}+1}=-n! \, \text{Li}_{n+1}\left(-e^{\mu/T}\right){,}
\end{equation}
and using the recurrent connections between the polylogarithms $\text{Li}_{n}(x)\,$
and
 $\text{Li}_{n}\,(x^{-1})\,$, one obtains the following expression:
 \begin{eqnarray}
\Delta E^{\nu_{e}} &=& \sqrt{2}
G_{F}\bigg[C_{V}^{e}\frac{\mu}{3\pi^2}\left(\mu^2+\pi^2
T^2\right)-\frac{2}{3\pi^2}\frac{E}{m_{W}^2}\left(\mu^4+2\pi^2\mu^2
T^2+\frac{7\pi^4}{15}T^4\right)
\nonumber\\[3mm]
&+& \frac{8}{5\pi^2}\frac{E^2
\mu}{m_{W}^4}\left(\mu^4+\frac{10\pi^2}{3}\mu^2
T^2+\frac{7\pi^4}{3}T^4\right)
\nonumber\\[3mm]
&-& \frac{64}{15\pi^2}\frac{E^3}{m_{W}^6}\left(\mu^6+5\pi^2\mu^4
T^2+7\mu^2\pi^4 T^4+\frac{31}{21}\pi^6 T^6\right)+\ldots\bigg]{.}
\label{eq:27}
\end{eqnarray}

It is worthwhile to note that the similar expression can be
written for the electron antineutrino. To write it down one has to make a
change $E \to - E$ in Eq.~\eqref{eq:27}. In Fig. 3, the additional electron neutrino energy
$\Delta E$ is illustrated as a function of the initial neutrino energy $E$. 
It is demonstrated that taking account of only few
terms in the series by the initial energy leads to an overestimation or
understatement of the additional energy.

\begin{figure}
\begin{center}
\includegraphics*[width=0.85\textwidth]{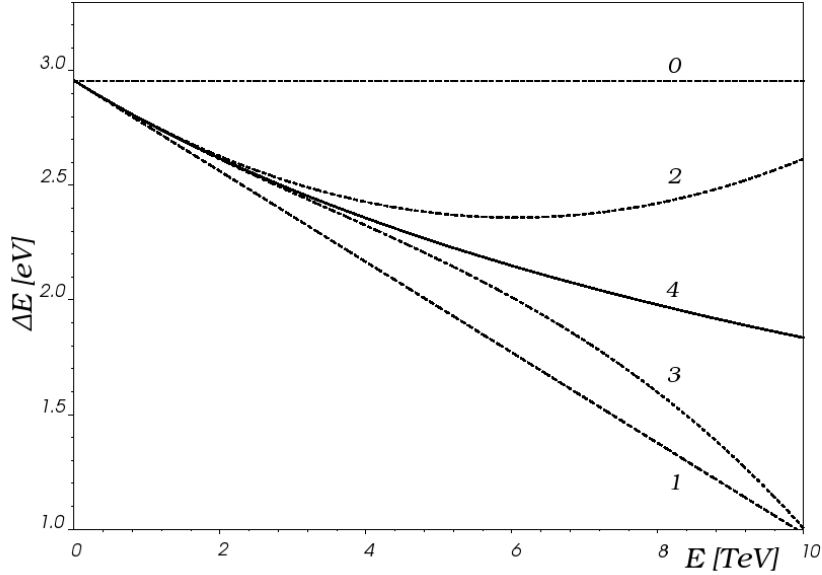}
\caption{Additional electron neutrino energy in the electron-positron
medium ($\mu_e \simeq 160$ MeV, $T \simeq 30$ MeV) as an expansion into the series by initial neutrino energy: 
{\it 0} is the local contribution; {\it 1}, {\it 2} and {\it 3} -- with consecutive adding 
of non-local terms $\sim E$, $\sim E^2$ and $\sim E^3$; {\it 4} is the exact function.}
\end{center}
\label{fig3}
\end{figure}

For a numerical estimation of the borders of the kinemetically
possible region for the $SL\nu$ process in a general case with using 
of Eq.~\eqref{eq:addl_E_SN}, let us take 
$\mu_{e}\simeq 160\,$MeV, $\mu_{\nu}\simeq \mu_{e}/4\simeq 40\,$MeV, see
e.g. Refs.~\refcite{Janka:2007} and~\refcite{Kitaura:2006}. 
The analysis displays that the process is forbidden for neutrinos of all flavors. For all 
types of antineutrinos the effect becomes 
possible for energies greater than $2 \times 10^{4}\,$ GeV.

As in the considered cases of ``cold'' plasma and of the neutron star interior, 
for electron neutrinos and antineutrinos the processes of the $W$--boson 
production on plasma electrons and positrons, 
$\nu_e + e^+ \to W^+$ and $\bar\nu_e + e^- \to W^-$, are dominating.
Using Eqs.~\eqref{eq:11}, \eqref{eq:width_via_DE}, one obtains the width of the process 
in the conditions of a hot dense plasma, $\mu_e \sim T \gg m_e$, for high neutrino energies, 
$E \gg m_W\, \Gamma_W/\mu_e$:
\begin{equation}
\label{eq:width_SN_W-} 
w (\bar\nu_e + e^- \to W^-) = \frac{G_{\mathrm{F}} m_W^4 T}{2 \sqrt{2} \, \pi \, E^2} 
\ln \left[ 1 + \exp \left( \frac{4 \mu_e E - m_W^2}{4 E T} \right) \right] \,.
\end{equation}
Taking here the limit of cold plasma, $T \to 0$, one readily comes to Eq.~\eqref{eq:width_NS}. 
The width of the $W^+$ production by $\nu_e$ on positrons can be obtained from Eq.~\eqref{eq:width_SN_W-} by 
the replacement $\mu_e \to - \mu_e$. 

Since in a dense plasma of the supernova core thermal neutrinos and antineutrinos of all flavors present, 
the processes of the $Z$--boson production should be also considered for the sake of completeness. 
Using Eqs.~\eqref{eq:11}, \eqref{eq:width_via_DE}, one obtains the width of the process where 
a high-energy antineutrino of the flavor $\ell$ scatters off a thermal $\nu_{\ell}$:
\begin{equation}
\label{eq:width_SN_Z} 
w (\bar\nu_{\ell} + \nu_{\ell} \to Z) = \frac{G_{\mathrm{F}} m_Z^4 T}{4 \sqrt{2} \, \pi \, E^2} 
\ln \left[ 1 + \exp \left( \frac{4 \mu_{\nu_{\ell}} E - m_Z^2}{4 E T} \right) \right] \,.
\end{equation}
The width of the process with a high-energy neutrino and a thermal antineutrino 
can be obtained from Eq.~\eqref{eq:width_SN_Z} by 
the replacement $\mu_{\nu_{\ell}} \to - \mu_{\nu_{\ell}}$. 
It should be noted that in the supernova core conditions $\mu_{\nu_{\ell}} \simeq 0$ for $\ell = \mu, \,\tau$. 

\section{Conclusion}
\label{sec:Conclusion}
We reexamine the previous results~\cite{Kuznetsov:2006,Kuznetsov:2007ijmpa} on a possibility
of the neutrino radiative conversion effect $\nu_L \to \nu_R + \gamma$
(``spin light of neutrino'', $SL \nu$) based on the additional neutrino energy in plasma, 
obtained in the local limit of the weak interaction (Wolfenstein energy) 
and with the first non-local correction. 
In the listed papers it was particularly demonstrated
that the possibility of the $SL \nu$ existence~\cite{Studenikin:2008} is overstated and 
the process is kinematically forbidden in almost all real
astrophysical conditions. The only question remained open whether
this effect is possible in the case of ultra-high neutrino
energies. In the present paper we eliminate this gap. Formulas for the 
neutrino and antineutrino additional energies in
plasma are obtained, based on the $W$-- and $Z$--boson propagators depending on the momentum transferred.
It should be noted that the question about any
observational realization of the studied process requires a
separate consideration.
For high energy neutrinos and antineutrinos, the processes of the $W$-- and $Z$--boson 
production on plasma, 
$\nu_e + e^+ \to W^+$, $\bar\nu_e + e^- \to W^-$ and $\bar\nu_{\ell} + \nu_{\ell} \to Z$, are dominating.

\section*{Acknowledgments}

This work was performed in the framework of realization of the
Federal Target Program ``Scientific and Pedagogic Personnel of the
Innovation Russia'' for 2009\,--\,2013 (State contract no. P2323)
and was supported in part by the Ministry of Education and Science
of the Russian Federation under the Program ``Development of the
Scientific Potential of the Higher Education'' (project no.
2.1.1/13011), and by the Russian Foundation for Basic Research
(project no. 11-02-00394-a).


\begin{thebibliography}{00}  

\bibitem{Wolfenstein:1978} L.~Wolfenstein, \textit{Phys.~Rev.~D} \textbf{17}, 9 (1978).

\bibitem{Studenikin:2006} A.~Studenikin, \textit{J.~Phys.~A: Math.~Gen.}, \textbf{39},
6769 (2006).

\bibitem{Kuznetsov:2006} A.\,V.~Kuznetsov and N.\,V.~Mikheev, \textit{Mod.~Phys.~Lett.~A}, \textbf{21}, 1769 (2006).

\bibitem{Kuznetsov:2007ijmpa} A.\,V.~Kuznetsov and N.\,V.~Mikheev, \textit{Int.~J.~Mod.~Phys.~A}, \textbf{22}, 3211 (2007).

\bibitem{Studenikin:2008} A.~Studenikin,
\textit{J.~Phys.~A: Math.~Gen.}, \textbf{41}, 164047 (2008).

\bibitem{Notzold:1988_NPB} D.~N\"otzold and G.~Raffelt,
\textit{Nucl.~Phys.~B}, \textbf{307}, 924 (1988).

\bibitem{Langacker:1992}  P.~Langacker and J.~Liu, \textit{Phys.~Rev.~D}, \textbf{46}, 4140 (1992).

\bibitem{Elmfors:1996}
P.~Elmfors, D.~Grasso and G.~Raffelt, \textit{Nucl.~Phys.~B}, \textbf{479}, 3 (1996).

\bibitem{Lunardini:2000}
C.~Lunardini and A.\,Yu~Smirnov, \textit{Nucl.~Phys.~B},
\textbf{583}, 260 (2000).

\bibitem{Lunardini:2001} 
C.~Lunardini and A.\,Yu~Smirnov, \textit{Phys.~Rev.~D}, \textbf{64}, 073006 (2001).

\bibitem{Sahu:2008}
S.~Sahu and W.\,-Y.\,P.~Hwang, \textit{Eur.~Phys.~J.~C}, \textbf{58}, 609 (2008).

\bibitem{Bjorken:1964}
   J.\,D.~Bjorken and S.\,D.~Drell,
   \textit{Relativistic Quantum Mechanics}
   (McGraw-Hill, New York, 1964). 

\bibitem{Potehin:2010}
A.\,Yu.~Potekhin, \textit{Usp. Fiz. Nauk} \textbf{180}, 1279 (2010) 
[\textit{Physics--Uspekhi} \textbf{53}, 1235 (2010)].

\bibitem{Janka:2007} H.\,-Th.~Janka, K.~Langanke and A.~Marek \textit{et al.},
\textit{Phys.~Rept.}, \textbf{442}, 38 (2007).

\bibitem{Kitaura:2006} F.\,S.~Kitaura, H.\,-Th.~Janka and
W.~Hillebrandt, \textit{Astron. Astrophys.}, \textbf{450}, 345, (2006).

\end{thebibliography}
\end{document}